\documentclass{article}
\usepackage{amsfonts}
\usepackage{amssymb}
\usepackage{amsbsy}
\usepackage{amsthm}
\newtheorem{lemma}{Lemma}
\newtheorem{theorem}[lemma]{Theorem}
\newtheorem{defin}[lemma]{Definition}

\newcommand{\bibtitle}[1]{\textit{#1}}
\newcommand{\bibseries}[1]{\textbf{#1}}
\newcommand{\bibyear}[1]{(#1)}

\newcommand{\setr}{\ensuremath{\mathbb{R}}}
\newcommand{\setc}{\ensuremath{\mathbb{C}}}
\newcommand{\supp}{\ensuremath{\mathrm{supp}\,}}
\newcommand{\cc}{\ensuremath{\mathrm{c.c.}\,}}
\newcommand{\ioo}[1]{\ensuremath{\:\left]#1\right[\:}}
\newcommand{\sL}{\ensuremath{\mathcal{L}}}
\newcommand{\sS}{\ensuremath{\mathcal{S}}}

\newcommand{\del}{\ensuremath{\nabla}}
\newcommand{\tdel}{\ensuremath{\tilde{\nabla}}}
\newcommand{\ndel}{\ensuremath{{\del^0}}}
\newcommand{\tG}{\ensuremath{\tilde{G}}}
\newcommand{\tR}{\ensuremath{\tilde{R}}}
\newcommand{\tM}{\ensuremath{\tilde{M}}}
\newcommand{\tg}{\ensuremath{\tilde{g}}}
\newcommand{\hR}{\ensuremath{\hat{R}}}

\newcommand{\hE}{\ensuremath{\hat{E}}}

\newcommand{\eT}{\ensuremath{\boldsymbol{T}}}
\newcommand{\eS}{\ensuremath{\boldsymbol{S}}}
\newcommand{\ej}{\ensuremath{\boldsymbol{j}}}
\newcommand{\erho}{\ensuremath{\boldsymbol{\rho}}}
\newcommand{\heS}{\ensuremath{\boldsymbol{\hat{S}}{}}}
\newcommand{\dt}{\ensuremath{\partial_t}}

\newcommand{\tr}{\ensuremath{\mathrm{tr}\,}}
\newcommand{\dal}{\ensuremath{\langle \! \langle}}
\newcommand{\dar}{\ensuremath{\rangle \! \rangle}}
\newcommand{\glx}{\ensuremath{\dal X \dar}}
\newcommand{\dsl}{\ensuremath{[ \! [}}
\newcommand{\dsr}{\ensuremath{] \! ]}} 
\newcommand{\px}{\ensuremath{\dsl X \dsr}}
\newcommand{\py}{\ensuremath{\langle Y \rangle}}

\newcommand{\ud}[2]{^#1_{\phantom{#1}#2}}
\newcommand{\du}[2]{_#1^{\phantom{#1}#2}}

\begin{document}

%
%

\author{Roger Bieli\\
Max Planck Institute for Gravitational Physics\\
Am M\"uhlenberg 1\\
14476 Golm\\
Germany
}

\title{Algebraic expansions for curvature coupled scalar field models}
\date{}

\maketitle

\begin{abstract}
 A late time asymptotic perturbative analysis of curvature coupled complex scalar field models
 with accelerated cosmological expansion is carried out on the level of formal power series
 expansions. For this, algebraic analogues of the Einstein scalar field equations in Gaussian
 coordinates for space-time dimensions greater than two are postulated and formal solutions
 are constructed inductively and shown to be unique. The results obtained this way are found
 to be consistent with already known facts on the asymptotics of such models.  In addition,
 the algebraic expansions are used to provide a prospect of the large time behaviour that might
 be expected of the considered models.
\end{abstract}

%
%

\section{Introduction}

A common and also remarkably consistent interpretation of current observational data on type
Ia supernovae distances, cosmic microwave background anisotropies and large scale structure
surveys is that the universe is of spatially flat Friedmann Robertson Walker type and is
currently undergoing accelerated expansion. However, the supposition of such a model opens a
gap between the predicted total energy density required by a flat space, the critical density,
and the sum of observed baryonic and dark matter density. An unknown, smooth component of
matter with negative pressure, dubbed \emph{dark energy}, is usually introduced to fill this
gap of about two thirds of the critical density.  For a recent review on dark matter and
dark energy see e.g. \cite{Sah04}. It is noteworthy that there are also attempts to avert
the need of dark energy at all, for instance by assuming Van der Waals type \cite{Cap02}
or viscous \cite{Fab05} instead of perfect matter fluids.

Trying to fathom the nature of dark energy, myriads of models have been brought up for it,
from the simplest, as just a positive cosmological constant \cite{Sah00} yielding an equation
of state parameter $w = -1$, to far richer, dynamical ones, where $w$ can vary in time. Many
of these models make use of various scalar fields to describe dark energy, although there
are not less successful approaches like the generalized Chaplygin gas \cite{Kam01} which
consider fluids that are to obey specific nonlinear equations of state. Models with real scalar
fields minimally coupled to the Ricci curvature and propagating in different (non-negative)
potentials, as proposed in \cite{Rat88}, are known as quintessence \cite{Cal98}, and cause
$w$ lying in the ``ordinary'' range between minus one and one.  If the kinetic energy of
such a field is sufficiently small compared with the potential energy, $w$ will be less than
$-1/3$, so driving the acceleration of the universe \cite{Ren0403,Ren0501}. An interesting
generalization of this is spintessence \cite{Boy02}, where the scalar field is assumed to be
complex \cite{GuH01} or even to possess an internal ${\rm O}(N)$ symmetry \cite{LiH01}. The
restrictions on quintessence can be relaxed in other ways by allowing arbitrary coupling to the
scalar curvature \cite{Far03}, to the matter fluid \cite{Ame00}, or, instead of a potential,
introducing non-standard kinetic terms \cite{Arm00}. The latter is known as $k$-essence. Doing
so enables $w$ to take values below $-1$, a case which has been discussed under the term
\emph{phantom energy} \cite{Cal02}. Phantom cosmologies are capable of exhibiting a certain
type of future singularity \cite{Cal02,Sta00}, the so called \emph{big rip} \cite{Cal03},
where the scale factor diverges in finite time.

A way of getting information about the late time asymptotics of cosmological models is to
consider formal power series expansions of their solutions as suggested by Starobinsky in
\cite{Sta83} and worked out in full detail by Rendall \cite{Ren0312} for the vacuum case.
Recently, for example, Heinzle and Rendall \cite{Hei05} calculated formal asymptotic expansions
for quintessence in an exponential potential, whereas in \cite{Bal05} and \cite{Gio05}
formal series were used to investigate the big rip singularity due to a phantom scalar field
and phantom barytropic perfect fluid respectively. The method presented in \cite{Ren0312}
encompasses two parts, a first algebraic one, in which the Einstein equations are used to
obtain algebraic equations for the series' coefficients that are then solved inductively,
and a second analytic one, where it is shown that those formal series indeed approximate
the cosmological solution to arbitrary high order for large times. Rendall \cite{Ren0312}
did also carry out the algebraic part for perfect fluids with linear equation of state,
more specifically for constant non-negative $w$ less than one. It is the aim of the present
paper to do such a perturbative analysis for curvature coupled complex scalar fields. For
this, section \ref{motivation} recapitulates the Einstein scalar field equations in $n+1$
decomposition with respect to a Gaussian coordinate system. In section \ref{setup} the formal
series are defined as elements of the algebra of generalized finite formal Laurent series over
a suitable ring. Section \ref{esfg} then gives partial and covariant derivatives as well as
(matrix) inverses of such series a precise meaning and postulates algebraic analogues of the
Einstein and scalar field equations obtained before. Finally, in section \ref{construct},
existence and uniqueness of solutions to this algebraic equations are shown.

%
%

\section{Motivation of the equations} \label{motivation}

In the algebraic treatment of scalar field models in the sections following, a set of
equations for formal series will be postulated. These equations shall be motivated here as
originating from a $n+1$ decomposition of the dynamics in Gaussian coordinates. For this,
let $\tg$ be a Lorentzian metric on $\tM := M \times I$, where $M$ is a smooth manifold of
dimension $n \geq 2$ and $I$ an open non-empty interval within the real numbers $\setr$,
such that the $n+1$ decomposition has unit lapse and vanishing shift, i.e. \[\tg| U \times
I = g_{ab} dx^a \otimes dx^b - dt \otimes dt\] for every chart\footnote{Compositions with
the natural projection $U \times I \to U$ are not written out explicitly.} $(U,x)$ of $M$,
where $t$ denotes the natural projection on $I$ and $g(\cdot,t_0)$ is the Riemannian metric
induced by $\tg$ on the hypersurface $t=t_0 \in I$. Then the Einstein equation $\tG = T$
on $\tM$ is equivalent to a system consisting of an evolution equation
\begin{equation} \label{evol.k}
 \dt k\ud{a}{b} = R\ud{a}{b} + (\tr k) k\ud{a}{b} - S\ud{a}{b} + \frac{\tr S - \rho} {n
  - 1} \delta\ud{a}{b} 
\end{equation}
as well as both an energy and a momentum constraint equation
\begin{eqnarray}
 R - k_{ij} k^{ij} + (\tr k)^2 & = & 2 \rho \label{constr.ham} \\
 \del_i k\ud{i}{a} - \del_a \tr k & = & j_a \label{constr.mom}.
\end{eqnarray}
Here, $k_{ab}$ is the scalar second fundamental form of the $t$-hypersurfaces in $\tM$,
which satisfies
\begin{equation} \label{evol.g}
 \dt g_{ab} = -2 k_{ab},
\end{equation}
whereas $R_{ab}$ and $R$ are the Ricci tensor and scalar curvature of $g$.  Moreover,
the following projections of the energy momentum tensor $T^{\mu \nu}$ with respect to the
$t$-hypersurfaces were used
\cite{Ren0312}:
\begin{equation} \label{emtproj}
 S^{ab}:=T^{ab};\qquad j^a:=T^{an}; \qquad \rho:=T^{nn}.
\end{equation}
Latin indices shall indicate spatial components and run from $0$ to $n-1$, whereas Greek indices
may cover the entire range, from $0$ to $n$.

If $\phi$ is a complex scalar field on $\tM$ coupled to the scalar curvature $\tR$ with strength
$\xi \in \setr$ and propagating in a smooth real potential $V$, then it fulfills the wave equation
\begin{equation} \label{sfg.n+1}
 \Box \phi - \xi \tR \phi - 2 \phi V'(\phi^\ast \phi) = 0
\end{equation}
and its energy momentum tensor is given by\footnote{ A recent and well-referenced discussion
on non-minimal coupling can be found in \cite{Far01}.}
\begin{eqnarray} \label{emtsf.n+1}
 T_{\mu \nu} & = & \frac{1}{2} \bigg[ (1-2\xi) \tdel_\mu \phi^\ast \tdel_\nu \phi + \left( 2\xi - \frac{1}{2}
  \right) g_{\mu \nu} \tdel_\alpha \phi^\ast \tdel^\alpha \phi - 2\xi \phi^\ast \tdel_\mu
  \tdel_\nu \phi \nonumber \\
 & & {} + 2\xi g_{\mu \nu} \phi^\ast \tdel_\alpha \tdel^\alpha \phi + \xi \tG_{\mu \nu} \phi^\ast
  \phi - g_{\mu \nu} V(\phi^\ast \phi) + \cc \bigg] .
\end{eqnarray}
The abbreviations $\Box$ for the covariant Laplacian $\tilde{\Delta}=\tdel_\alpha \tdel^\alpha$
of $\tM$ as well as $\cc$ for the complex conjugate of all the preceding terms were used. Note
that in contrast to real scalar fields, where the potential term is usually written as $V(\phi)$,
it is here defined to be $V(\phi^\ast \phi)$ for having $V$ a function of a real variable. The
explicit occurrence of the Einstein tensor in (\ref{emtsf.n+1}) can be circumvented by assuming
$1 - \xi \phi^\ast \phi > 0$ and introducing an effective energy momentum tensor \cite{Bar92}
\begin{equation} \label{effemt}
 \eT := \frac{T - \xi \tG \phi^\ast \phi}{1 - \xi \phi^\ast \phi}.
\end{equation}
The wave equation (\ref{sfg.n+1}) is satisfied if and only if
\begin{equation} \label{sfg.n}
 \Delta \phi + (\tr k) \dt \phi - \dt^2 \phi - \xi \tR \phi - 2 \phi V'(\phi^\ast \phi) = 0
\end{equation}
holds on $\tM$, where the ambient scalar curvature $\tR$ can be written as
\begin{equation} \label{tR}
 \tR = R - 2\dt \tr k + k_{ij}k^{ij} + (\tr k)^2.
\end{equation}
The Einstein equations are then equivalent to $\tG = \eT$ which in turn is valid exactly if the
(effective) evolution and constraint quantities \cite{Ren0312}
\begin{eqnarray}
 \hE\ud{a}{b} & := & \dt \sigma\ud{a}{b} - \left[ \hR\ud{a}{b} + (\tr k) \sigma\ud{a}{b} - 
  \heS\ud{a}{b} \right] \label{effevol.tl} \\
 E & := & \dt \tr k - \left[ R + (\tr k)^2 + \frac{1}{n-1} \tr \eS - \frac{n}{n-1} \erho 
  \right] \label{effevol.tr} \\
 C & := & R - k_{ij} k^{ij} + (\tr k)^2 - 2 \erho \label{effconstr.ham} \\
 C_a & := & \del_i k\ud{i}{a} - \del_a \tr k - \ej_a \label{effconstr.mom}
\end{eqnarray}
all vanish identically. In this paper, a hat shall denote the tensor's traceless part and
$\sigma:=\hat{k}$. The
projections of the effective energy momentum tensor $\eT$ can be written out as
\begin{eqnarray}
 \eS\ud{a}{b} & = & \frac{1}{2} (1-\xi \phi^\ast \phi)^{-1} \bigg[ (1-2\xi) \del^a \phi^\ast
  \del_b \phi - 2 \xi \phi^\ast \big[ \del^a \del_b \phi + k\ud{a}{b} \dt \phi \big]
  \nonumber \\
 & & {} + \bigg\{ \left( 2\xi - \frac{1}{2} \right) \big[ \del_i \phi^\ast \del^i \phi -
  \dt \phi^\ast \dt \phi \big] + 2 \xi \phi^\ast \big[ \Delta \phi \nonumber \\
 & & {} + (\tr k) \dt \phi - \dt^2 \phi \big] - V( \phi^\ast \phi) \bigg\}
  \delta\ud{a}{b} + \cc \bigg] \label{effS} \\
 \ej^a & = & \frac{1}{2} (1-\xi \phi^\ast \phi)^{-1} \bigg[ - (1-2\xi) \del^a \phi^\ast \dt \phi +
  2 \xi \phi^\ast \big[ \dt \del^a \phi \nonumber \\
 & & {} - k\ud{a}{i} \del^i \phi \big] + \cc \bigg] \label{effj} \\
 \erho & = & \frac{1}{2} (1-\xi \phi^\ast \phi)^{-1} \bigg[ \frac{1}{2} \dt \phi^\ast \dt \phi -
  \left( 2\xi - \frac{1}{2} \right) \del_i \phi^\ast \del^i \phi - 2 \xi \phi^\ast \big[
  \Delta \phi \nonumber \\
 & & {} + (\tr k) \dt \phi \big] + V(\phi^\ast \phi) + \cc \bigg]. \label{effrho}
\end{eqnarray}  

Finally, two useful consistency conditions can be obtained by differentiating the energy and
momentum constraint quantities (\ref{effconstr.ham}), (\ref{effconstr.mom}) with respect to time
\begin{eqnarray}
 \label{consist.ham} \dt C & = & 2 (\tr k) C - 2 \del^i C_i - 2 \sigma\ud{i}{j}
  \hE\ud{j}{i} + 2 \left( 1 - \frac{1}{n} \right) (\tr k) E \nonumber \\
  & & {} - 2 \big[ \dt \erho + \del_i \ej^i - k_{ij} \eS^{ij} - (\tr k)
  \erho \big] \\ 
 \label{consist.mom} \dt C_a & = & (\tr k) C_a - \frac{1}{2} \del_a C + \del_i
  \hE\ud{i}{a} - \left( 1 - \frac{1}{n} \right) \del_a E \nonumber \\
  & & {} - g_{ai} \big[ \dt \ej^i + \del_j \eS^{ij} - 2 k\ud{i}{j} \ej^j -
  (\tr k) \ej^i \big]
\end{eqnarray}
and identifying the matter terms in square brackets above as the temporal and spatial components of
the covariant divergence of the effective energy momentum tensor respectively.

%
%

\section{Algebraic setup} \label{setup}

To investigate the asymptotics of curvature coupled scalar field models algebraically, an
ansatz for both the metric and the matter field in terms of formal power series is made. Then,
the Einstein scalar field equations (\ref{evol.k}), (\ref{constr.ham}), (\ref{constr.mom})
and (\ref{sfg.n}) provide relations on the coefficients of those series. An ansatz found to
be suitable is of the form
\begin{displaymath} \label{ansatz}
 \sum_{m \in D} \sum_{s \in S_m} \sum_{l \in L_{m,s}} a_{m,s,l}(x) t^l e^{-(m+is) H t},
\end{displaymath}
where $D$ belongs to a certain class of subsets of the reals given below and $S_m$ and $L_{m,s}$
are finite sets of real and natural numbers respectively, this for $m \in D$ and every $s \in
S_m$. Furthermore, $H$ is a positive constant playing the role of an asymptotic value of the
Hubble parameter, and $a_{m,s,l}(x)$ is a coefficient independent of $t$. Non-vanishing terms
with $l \neq 0$ will be called \emph{logarithmic}, such with $s \neq 0$ \emph{oscillatory}.

In order to give the series above a precise meaning, it is now shown that they can be taken
as elements of a specific algebra providing the structure required. For this, let $R$ be
a complex algebra\footnote{Throughout this paper, an algebra is considered associative,
commutative and unitary.} and $R\glx$ the set of functions from $\setr$ to $R$ whose support
is discrete\footnote{A subset $A$ of a topological space $X$ is called discrete iff $A$
has no limit points in $X$.} and bounded from below. Then this set together with point-wise
addition, point-wise exterior multiplication and the Cauchy product forms the complex algebra of
generalized finite formal Laurent series in one variable $X$ over $R$, where the supposition
on the support of the functions guarantees that the sums in the inner multiplication are
all finite and therefore well-defined. Note that for complex $\kappa \in \setc$ and $f,g
\in R\glx$ the support of $\kappa f$ is a subset of the support of $f$, that of $f+g$ is a
subset of the union of the supports of $f$ and $g$, whereas the support of the product $fg$
lies within the point-wise sum of the supports of $f$ and $g$. As usual, $X^\alpha$ denotes
the map for which $X^\alpha(m)=1_R$, the identity in $R$, for $m=\alpha$ and zero otherwise.
By virtue of the identity $1_{R\glx}=X^0$ in $R\glx$ the algebra $R$ is identified with the
subalgebra $R 1_{R\glx}$ in $R\glx$. In addition, a function taking values in $R$ defined on
a subset of $\setr$ is always identified with its trivial extension, where zero is assigned
to points outside the function's domain. It will be useful to say a $f \in R\glx$ being of
order not less than $\alpha$, written $f = O(X^\alpha)$, iff $\alpha \leq \supp f$, being of
order greater than $\alpha$, written $f = o(X^\alpha)$, iff $\alpha < \supp f$, i.e. non-zero
coefficients occur not before or even after $\alpha \in \setr$ respectively.

There are three subalgebras of $R\glx$ which will be used further on. Firstly, the polynomials
$R[X]$ over $R$ encompassing exactly those series supported on a finite subset in the natural
numbers, secondly, the power series $R\px$ over $R$, obtained by  allowing series supported
on any set of naturals, and thirdly, the generalized polynomials $R\langle X \rangle$ over
$R$, whose support is a finite subset in the real numbers. With this, and by noting that
the set of smooth complex valued functions $C^\infty(M, \setc)$ on the $n$-manifold $M$
possesses the structure of a complex algebra, it is possible to define an algebra $\sL :=
C^\infty(M,\setc)[Z]\py\glx$ over $\setc$ in three variables $X$, $Y$ and $Z$ successively,
i.e. $\sL = ((C^\infty(M,\setc)[Z])\py)\glx$. The identification of these variables with distinct
elements of $\sL$ follows from the outline in the previous paragraph. An element of the algebra
$\sL$ can be regarded as a series of the form (\ref{ansatz}) when the variable $X$ plays the
role of the exponential factor $e^{-Ht}$, $Y$ that of the oscillatory factor $e^{-iHt}$ and $Z$
that of the logarithmic factor $t$. The coefficients are smooth complex functions on $M$. It
is worth remarking that this definition of the formal series is totally intrinsic to $M$,
the appearance of $t$ in (\ref{ansatz}) is merely symbolic and intended to suggest arithmetic
rules rather than being a reference to the ambient manifold $\tM$. To have the setup complete,
let $\sS_u$ denote the subalgebra of exactly those elements of $\sL$, which are of order not
less than zero and have neither oscillatory nor logarithmic terms occurring before $u \in
\setr$. Using the above identifications, the absence of those terms can also be expressed as
the restriction of the series to the open interval $\ioo{-\infty,u}$ lying in
$C^\infty(M,\setc)\glx$.

%
%

\section{Algebraic Einstein scalar field equations} \label{esfg}

After these preparations, it is possible to postulate Einstein and scalar field equations in
terms of formal series. For a $f \in \sL$, formal spatial derivatives shall be defined to act
point-wise, $( \partial_a f)_{m,s,l} := \partial_a (f)_{m,s,l}$, and the formal time derivative
according to $ (\dt f)_{m,s,l} := - (m+is) H (f)_{m,s,l} + (l+1) (f)_{m,s,l+1}$, $m,s,l \in
\setr$. This imitates the effect the partial derivatives would have when applied term-wise
to an ansatz of the form (\ref{ansatz}).  Here, the notation $(f)_{m,s,l}$ for $f(m)(s)(l)
\in C^\infty(M,\setc)$ is used. Those formal derivatives are derivations on the algebra $\sL$
commuting pairwise and leaving the subalgebras $\sS_u$, $u \in \setr$, invariant. To have
a notion of reality for formal series, define complex conjugation as a ring homomorphism $f
\mapsto f^\ast$ by $(f^\ast)_{m,s,l} := (f)^\ast_{m,-s,l}$ for all $m,s,l \in \setr$ and say
that $f$ is real iff $f^\ast = f$.

Fix a chart on $M$ which, for the sake of simplicity, may be global and let $(g_{ab})$ be a real
symmetric family in $\sS_u X^{-v}$ for some $u>0$ and $v \in \setr$ with $(g_{ab})_{-v,0,0}$
(the components of) a Riemannian metric. It can be shown that there exists exactly one real
symmetric family $(g^{ab})$ in $\sS_u X^v$ with $g_{ai} g^{ib} = \delta\du{a}{b}$ and $g^{ai}
g_{ib} = \delta\ud{a}{b}$. Thus, together with the formal partial derivatives above, algebraic
equivalents for the components of the Levi-Civita connection, the Ricci tensor and hence for
covariant derivatives and scalar curvature of $M$ can be obtained by setting
\begin{eqnarray*}
 \Gamma^c_{ab} & = & \frac{1}{2} g^{ic} ( \partial_a g_{ib} + \partial_b
  g_{ai} - \partial_i g_{ab} ) \ {\rm and} \\
 R_{ab} & = & \partial_i \Gamma^i_{ab} - \partial_a
  \Gamma^i_{ib} + \Gamma^i_{ac} \Gamma^i_{ij} - \Gamma^i_{aj}
  \Gamma^j_{ib}.
\end{eqnarray*}
In addition, equation (\ref{evol.g}) is used to define $k_{ab}$ algebraically and, in this context,
also intrinsically to $M$.

Having the formal series for the geometric quantities at hand, algebraic representations of
the components of the energy momentum tensor can be found. For this, let $\xi, \mu, \Lambda
\in \setr$ be arbitrary values for the coupling constant, field mass and cosmological constant
respectively and assume $H > 0$. A formal scalar field $\phi \in \sL$ of order higher than zero
and a smooth self interaction potential shall be given such that $V(0) = \Lambda$ and $V'(0)
= \mu^2 /2$, i.e. the first two coefficients of the McLaurin expansion may be denominated
explicitly.  A formal equivalent of $V(\phi^\ast \phi)$ is naturally obtained by applying the
substitution homomorphism induced by $\phi^\ast \phi$ to the McLaurin expansion of $V$,
hence setting
\begin{displaymath}
 W := \sum_{n=0}^\infty \frac{1}{n!} (\partial^n V)(0) (\phi^\ast \phi)^n
\end{displaymath}
which is well-defined because of $\phi^\ast \phi = o(1)$ and so the sum is point-wise finite.
Substitution into the expansion of $V'$ shall yield $W'$ in the same manner.  Also due to
$\phi^\ast \phi = o(1)$, the element $1-\xi \phi^\ast \phi$ has a multiplicative inverse in
$\sL$ which is exactly of order zero. Now the components of the effective energy momentum
tensor (\ref{effS}), (\ref{effj}) and (\ref{effrho}), the evolution and constraint quantities
(\ref{effevol.tl}), (\ref{effevol.tr}), (\ref{effconstr.ham}) and (\ref{effconstr.mom}) as well
as the scalar field equation (\ref{sfg.n}) make sense in terms of formal series and therefore
lead to the following definition \ref{alg.esfg}. It is convenient to introduce the abbreviations
\begin{eqnarray*}
 K & := & \left[ n^2/4 - \xi n(n+1) \right] H^2 \\
 k_1 & := & \left\{ \begin{array}{lll} n/2 & {\rm for} & \mu^2 > K \\ n/2 - \sqrt{K-\mu^2}/H &
  {\rm for} & \mu^2 \leq K \end{array} \right. \\
 k_2 & := & \left\{ \begin{array}{lll} n/2 & {\rm for} & \mu^2 > K \\ n/2 + \sqrt{K-\mu^2}/H &
  {\rm for} & \mu^2 \leq K \end{array} \right.
\end{eqnarray*} 
here.
\begin{defin} \label{alg.esfg}
 Let $(g_{ab})$ be a real symmetric family in $\sS_u X^{-v}$ for some $u>0$, $v \in \setr$, with
 $(g_{ab})_{-v,0,0}$ a Riemannian metric and $\phi \in \sL$ of order greater than zero. Then
 the pair $(g_{ab},\phi)$ is said to be a solution of the (algebraic) Einstein or scalar field
 equations at relative order $m \in \setr$ iff the evolution and constraint quantities vanish
 at $m$, so $(\hE\ud{a}{b},E,C,C_a)(m)=0$, or (\ref{sfg.n}) holds at $m+k_1$ respectively.
 It is called a solution of the Einstein or scalar field equations iff it is a solution of
 the corresponding equation at every relative order $m \in \setr$.
\end{defin} 
Furthermore, the concept of relative orders shall be extended to the metric and field by
defining the coefficients of $g_{ab}$ and $\phi$ at relative order $m$ to be $g_{ab}(m-2)$
and $\phi(m+k_1)$ respectively.

Simple but useful necessary conditions for the existence of algebraic solutions as well
as relations of the asymptotic Hubble constant $H$ or the principal frequency of field
oscillations to the cosmological constant $\Lambda$ or field mass $\mu$, respectively, can
be obtained by focusing attention to the lowest order terms in the series for both metric and
field. Let $(g_{ab},\phi)$ be a solution of the Einstein scalar field equations. Prescribing
the metric $g_{ab}$ to lie within $\sS_n X^{-2}$ and have coefficients $(g_{ab})_{-2,0,0}$
that make up a Riemannian metric, implies, as mentioned above, $g^{ab} \in \sS_n X^2$ which in
turn requires the components of the connection $\Gamma^c_{ab}$ and Ricci tensor $R_{ab}$ being
real and in $\sS_n$.  This in turn leads to $R\ud{a}{b}, R \in \sS_n X^2$ both real, whereas
equation (\ref{evol.g}) identifies $k_{ab} \in \sS_n X^{-2}$ real and so either $k\ud{a}{b}$
and $\tr k$ as real elements of $\sS_n$. The same equation (\ref{evol.g}) also determines
their lowest order coefficients entirely,
\begin{equation} \label{init.k}
 (k\ud{a}{b})_{0,0,0} = -H \delta\ud{a}{b} \,; \qquad (\sigma\ud{a}{b})_{0,0,0} = 0 \,;
 \qquad (\tr k)_{0,0,0} = -n H.
\end{equation} 

Further, assume the field $\phi$ being in $\sS_{k_2}$ and not equal to $0$, then the minimum
$r$ of the support of $\phi$ exists and the scalar field equation (\ref{sfg.n}) at $r$ reads
\begin{eqnarray*}
 \lefteqn{ (l+1)(l+2)(\phi)_{r,s,l+2} + (l+1) (n - 2r - 2is) H
 (\phi)_{r,s,l+1} } \\
 & & {} + \left\{ \mu^2 - s^2 H^2 + \left[ r(r-n) + \xi n(n+1)
 \right] H^2 + i s (2r-n) H^2 \right\} (\phi)_{r,s,l} = 0
\end{eqnarray*} 
for all $s,l \in \setr$. From this, it can be seen that $\phi$ is actually an element of
$\sS_{k_2-k_1} X^{k_1}$ and that the assertion
\begin{eqnarray*}
 \phi(r) \in C^\infty(M,\setc) & {\rm for} & \mu^2 < K \\
 \phi(r) \in C^\infty(M,\setc)+C^\infty(M,\setc)Z & {\rm for} & \mu^2 = K \\
 \phi(r) \in C^\infty(M,\setc)Y^{-\omega}+C^\infty(M,\setc)Y^\omega & {\rm for} & \mu^2 > K
\end{eqnarray*}
holds, i.e. for subcritical field masses there are neither logarithmic nor oscillatory
\emph{initial} terms, for critical field masses at most one logarithmic but no oscillatory
initial terms and for supercritical field masses no logarithmic but in general two oscillatory
initial terms at frequency $\omega H$ present, were
\begin{equation} \label{freq}
 \omega := \left\{ \begin{array}{lll}
  \sqrt{ \mu^2 - K } / H & {\rm for} & \mu^2 > K \\ 0 & {\rm for} & \mu^2 \leq K.
 \end{array} \right.  
\end{equation}
Using the lowest order coefficients for $k\ud{a}{b}$ known from (\ref{init.k}), the scalar field
equation (\ref{sfg.n}) can be written in a form suitable for calculating the field inductively,
namely for every positive relative order $m > 0$
in case of subcritical or critical field masses $\mu^2 \leq K$ as
\begin{eqnarray} \label{sfg.noc}
 \lefteqn{ (l+1)(l+2) (\phi)_{m+k_1,s,l+2} - (l+1) (2m+k_1-k_2+2is) H (\phi)_{m+k_1,s,l+1} }
  \nonumber \\
 & & {} + (m+is)(m+k_1-k_2+is) H^2 (\phi)_{m+k_1,s,l} = {\rm LOT}_{m+k_1,s,l},
  \hspace{0.1\textwidth}
\end{eqnarray}
in case of supercritical field masses $\mu^2 > K$ as
\begin{eqnarray} \label{sfg.oc}
 \lefteqn{ (l+1)(l+2) (\phi)_{m+k_1,s,l+2} -2(l+1)(m+is) H (\phi)_{m+k_1,s,l+1} } \nonumber \\
 & & {} + (m+is-i\omega)(m+is+i\omega) H^2 (\phi)_{m+k_1,s,l} = {\rm LOT}_{m+k_1,s,l},
  \hspace{0.1\textwidth}
\end{eqnarray}
where in ${\rm LOT}_{m+k_1,s,l}$ only field coefficients of relative order lower than $m$
and metric coefficients of relative order lower than or equal to $m$ occur.

From the fact that $\phi$ is in $\sS_{k_2-k_1} X^{k_1}$ it follows that $\phi^\ast \phi$,
$W$ and $W'$ lie within $\sS_n$, so do the components of the effective energy momentum tensor
$\eS\ud{a}{b}, \ej_a, \erho$ and with them, finally, all the evolution and constraint quantities
$E\ud{a}{b},E,C,C_a$. Furthermore, according to the equations (\ref{effS}), (\ref{effj}) and
(\ref{effrho}),
\begin{equation} \label{initT}
 (\eS\ud{a}{b})_{0,0,0} = - \Lambda \delta\ud{a}{b} \,; \qquad (\ej_a)_{0,0,0} = 0 \,;
 \qquad (\erho)_{0,0,0} = \Lambda
\end{equation}
is implied, and hence the validity of the Einstein equations to relative order zero requires in
particular $\Lambda > 0$ and
\begin{equation} \label{hubble}
 H = \sqrt{ \frac{2 \Lambda}{n(n-1)} }.
\end{equation}
Their validity up to but not including $k_0:=\min \{ 2, 2 k_1 \}$ yields the vanishing of
$g_{ab}$ and hence $k_{ab}$ on the open interval $\ioo{-2,k_0-2}$, of $g^{ab}$ on the interval
$\ioo{2,k_0+2}$, thus of $k\ud{a}{b}$, $\Gamma^c_{ab}$ and $R_{ab}$ on the interval $\ioo{0,k_0}$
and so of $R\ud{a}{b}$ on $\ioo{2,k_0+2}$ by induction.

Taking the last statement above into account, it is possible to express the evolution and
constraint quantities in a coefficient-wise manner for relative orders $m \ge k_0$ as
\begin{eqnarray}
 \label{egevol.tl} (\hE\ud{a}{b})_{m,s,l} & = & (n-m-is) H (\sigma\ud{a}{b})_{m,s,l} + (l+1)
  (\sigma\ud{a}{b})_{m,s,l+1} - (\hR\ud{a}{b})_{m,s,l} \nonumber \\ 
  & & {} - \sum_{p+q=m} \sum_{u+v=s} \sum_{\kappa+\lambda=l} (\sigma\ud{a}{b})_{p,u,\kappa} 
  (\tr k)_{q,v,\lambda} + (\heS\ud{a}{b})_{m,s,l} \\
 \label{egevol.tr} (E)_{m,s,l} & = & (2n-m-is) H (\tr k)_{m,s,l} + (l+1) (\tr
  k)_{m,s,l+1} - (R)_{m,s,l} \nonumber \\
  & & {} - \sum_{p+q=m} \sum_{u+v=s} \sum_{\kappa+\lambda=l} (\tr k)_{p,u,\kappa} (\tr
  k)_{q,v,\lambda} \nonumber \\
  & & {} - \left[ (\tr \eS)_{m,s,l} + n (\erho)_{m,s,l} \right] / (n-1) \\
 \label{egconstr.ham} (C)_{m,s,l} & = & -(2n-2) H (\tr k)_{m,s,l} + (R)_{m,s,l} 
  - 2 (\erho)_{m,s,l} \nonumber \\
  & & {} + \sum_{p+q=m} \sum_{u+v=s} \sum_{\kappa+\lambda=l} \Big[ - (k\ud{i}{j})_{p,u,
  \kappa} (k\ud{j}{i})_{q,v,\lambda} \nonumber \\
  & & {} + (\tr k)_{p,u,\kappa} (\tr k)_{q,v, \lambda} \Big] \\
 \label{egconstr.mom} (C_a)_{m,s,l} & = & \ndel_i (k\ud{i}{a})_{m,s,l} - \ndel_a (\tr
  k)_{m,s,l} - (\ej_a)_{m,s,l} \nonumber \\
  & & {} - \sum_{p+q=m} \sum_{u+v=s} \sum_{\kappa+\lambda=l} \Big[
  (\Gamma^i_{aj})_{p,u,\kappa} (k\ud{j}{i})_{q,v,\lambda} \nonumber \\
  & & {} - (\Gamma^i_{ij})_{p,u,\kappa} (k\ud{j}{a})_{q,v,\lambda} \Big] ,
\end{eqnarray} 
where the indices $p$ and $q$ are never less than $k_0$ and $\ndel$ denotes the Levi-Civita
connection induced on $M$ by the Riemannian metric $(g_{ab})_{-2,0,0}$. For the same relative
orders, equation (\ref{evol.g}) reads
\begin{eqnarray} \label{cgevol.g}
 \lefteqn{ (m+is) H (g_{ab})_{m-2,s,l} - (l+1) (g_{ab})_{m-2,s,l+1} = } \nonumber \\
 & & 2 (g_{ai})_{-2,0,0} (k\ud{i}{b})_{m,s,l} + 2 \sum_{p+q=m} \sum_{u+v=s} 
 \sum_{\kappa+\lambda=l} (g_{ai})_{p-2,u,\kappa} (k\ud{i}{b})_{q,v,\lambda}.
\end{eqnarray} 

Assuming validity of the Einstein scalar field equations up to but not including relative order $m
\geq k_0$, the consistency conditions (\ref{consist.ham}) and (\ref{consist.mom}) simplify to
\begin{equation} \label{cc.ham}
 (2n-m-is) H (C)_{m,s,l} + (l+1) (C)_{m,s,l+1} = -2 (n-1) H (E)_{m,s,l}
\end{equation}
and
\begin{eqnarray} \label{cc.mom}
 \lefteqn{ (n-m-is) H (C_a)_{m,s,l} + (l+1) (C_a)_{m,s,l+1} = } \nonumber \\
 & & - \frac{1}{2} \del_a (C)_{m,s,l} + \ndel_i (\hE\ud{i}{a})_{m,s,l} - \left(
  1 - \frac{1}{n} \right) \del_a (E)_{m,s,l}.
\end{eqnarray}

%
%

\section{Construction of the solution} \label{construct}

Using the relations for the coefficients of metric and scalar field derived in the previous
section, existence and uniqueness of algebraic solutions of the Einstein scalar field equations
can be shown by (transfinite) induction. After making some considerations about the support of
the required formal series, \emph{initial solutions} are constructed up to relative order $n$,
exclusive, which by themselves will allow stating a necessary and sufficient condition for
the existence of solutions up to beyond relative order $n$. For this, let a coupling constant
$\xi \in \setr$, a cosmological constant $\Lambda > 0$, a field mass $\mu \in \setr$ and a
smooth real valued self interaction potential $V$, defined on an interval containing zero,
with $V(0)=\Lambda$, $V'(0)=\mu^2/2$ be given, define $H$ and $\omega$ by (\ref{hubble}) and
(\ref{freq}) respectively and assume $\mu^2 > - \xi n (n+1) H^2$. The condition on the field
mass ensures that $k_1$ is positive and therefore the constructed field will be of order
greater than zero.

Let $D$ be the smallest $+$-stable discrete non-negative set containing $0, 1, k_1$ and $k_2
- k_1$, where $+$-stable means the set is invariant under point-wise addition, i.e. $D+D=D$.
For arbitrary $g_{ab}$ and $\phi$ both in $\sL$ suppose $(g_{ab})_{-2,0,0}$ to be a Riemannian
metric, the support of $g_{ab}$ to be contained in $D-2$ and that of $\phi$ to be contained in
$D+k_1$. Then, the support of $k_{ab}$ is a subset of $D-2$, the support of $g^{ab}$ one of $D+2$
and so the supports of all the terms in the components of the effective energy momentum tensor
(\ref{effemt}), in the evolution and constraint quantities (\ref{effevol.tl}), (\ref{effevol.tr}),
(\ref{effconstr.ham}), (\ref{effconstr.mom}) and hence of those quantities themselves lie within
$D$. The same is true for the left hand side of the scalar field equation (\ref{sfg.n}). This
means that it suffices to verify the validity of the Einstein scalar field equations on the
set $D$. Since $D$ is well-ordered by the induced order of $\setr$, transfinite induction is
available immediately. Of course, because $D$ is at most countable, ordinary induction could
also be used after identifying $D$ with its isomorphic ordinal.

As mentioned above, existence and uniqueness of \emph{initial solutions} will now be proved. More
precisely:
\begin{lemma} \label{initsol}
 Let $A_{ab} \in C^\infty(M)$ be the components of a Riemannian metric and $\phi_0, \phi_1
 \in C^\infty(M,\setc)$ smooth complex valued functions on $M$. Then there exists a solution
 $(g_{ab}, \phi)$ of the Einstein scalar field equations up to relative order $n \geq 2$,
 exclusive, with $g_{ab} \in \sS_n X^{-2}$, $\phi \in \sS_{k_2-k_1} X^{k_1}$ such that
 \begin{displaymath}
  \begin{array}{llrcl}
   {\rm (G1)} & \hfill & g_{ab}(-2) & = & A_{ab} \\[1ex]
   {\rm (P1)} & & \phi(k_1) & = & \left\{
   \begin{array}{lll}
    \phi_0 & {\rm for} & \mu^2 < K \\
    \phi_0 + \phi_1 Z & {\rm for} & \mu^2 = K \\
    \phi_0 Y^{-\omega} + \phi_1 Y^\omega & {\rm for} & \mu^2 > K
   \end{array} \right. \\[4ex]
   {\rm (P2)} & & \lefteqn{ \phi(k_2) \in \phi_1 + C^\infty(M,\setc)Z \ {\rm for} \ \mu^2
   < K } \hfill
  \end{array}
 \end{displaymath}
 are valid. The solution is unique up to, but not including, relative order $n$.
\end{lemma}

\begin{proof}
 The construction is done by induction over relative orders in the set $D \cap \ioo{-\infty,n}$
 using the following argument: For a $0 \leq m < n$ let $g_{ab} : \ioo{-\infty,m-2} \to
 C^\infty(M)$ and $\phi : \ioo{-\infty,m+k_1} \to C^\infty(M,\setc)[Z]\py$ be constructed
 such that $(g_{ab})$ is symmetric, the Einstein scalar field equations are satisfied up
 to relative order $m$, exclusive, and that for positive $m$ the conditions (G1), (P1),
 for $m+k_1 > k_2$ additionally (P2), hold. Moreover, the support of $g_{ab}$ may be finite,
 having exactly the point $-2$ lying below $k_0-2$, the support of $\phi$ be finite as well,
 having no points lying below $k_1$. Finally, for all $t$ less than both $m+k_1$ and $k_2$,
 $\phi(t)$ may belong to the subalgebra $C^\infty(M,\setc)$. Then the maps are extended to
 relative order $m$ dependent on the value of $m$:
 \begin{description}
  \item[$m=0:$]
   Set $g_{ab}(-2):=A_{ab}$, $\phi(k_1):=\phi_0$ for $\mu^2>K$, $\phi(k_1):=\phi_0 + \phi_1 Z$
   for $\mu^2=K$ and $\phi(k_1):=\phi_0 Y^{-\omega} + \phi_1 Y^\omega$ for $\mu^2>K$, then
   (G1) and (P1) are satisfied, the Einstein equations hold at relative order $0$ due to the
   choice of $H$, the scalar field equation holds there due to the choices of $\omega$ and $k_1$.
  \item[$0<m<k_0:$]
   Setting $g_{ab}(m-2):=0$ fulfills the Einstein equations at relative order $m$ because
   $k_0$ is less than or equal to both $2 k_1$ and $2$.  If $\mu^2 > K$ or $\mu^2 \leq K$
   and $m+k_1 \neq k_2$, the relations (\ref{sfg.oc}) or (\ref{sfg.noc}) yield a $\phi(m+k_1)$
   in $C^\infty(M,\setc)[Z]\py$ such that the scalar field equation is satisfied at relative
   order $m$. (The occurring linear systems can, for example, be solved successively by
   starting at sufficiently large values of $l$ and descending towards zero.) For $m+k_1 <
   k_2$ $\phi(m+k_1)$ actually lies within $C^\infty(M,\setc)$. If, on the other hand, $\mu^2
   \leq K$ and $m+k_1 = k_2$, then relation (\ref{sfg.noc}) implies the existence of a $\chi
   \in C^\infty(M,\setc)$ such that the scalar field equation is solved at relative order $m$
   by setting $\phi(m+k_1):=\phi_0 + \chi Z$. This, in particular, suffices (P2).
  \item[$k_0 \leq m < n:$]
   Using equations (\ref{egevol.tl}), (\ref{egevol.tr}) and (\ref{cgevol.g}), $g_{ab}(m-2)
   \in C^\infty(M)$ are found such that $(g_{ab})$ is symmetric and cancels the evolution
   quantities $\hE\ud{a}{b}(m)$ and $E(m)$ at relative order $m$. Now the consistency condition
   (\ref{cc.ham}) shows that the Hamiltonian constraint quantity $C(m)$ at relative order
   $m$ vanishes and (\ref{cc.mom}) subsequently implies the same for the momentum constraint
   quantities $C_a(m)$. Thus, the Einstein equations are solved at relative order $m$. The
   extension of $\phi$ can be carried out exactly as in the previous case: If $\mu^2 > K$
   or $\mu^2 \leq K$ and $m+k_1 \neq k_2$, the relations (\ref{sfg.oc}) or (\ref{sfg.noc})
   yield a $\phi(m+k_1)$ in $C^\infty(M,\setc)[Z]\py$ such that the scalar field equation
   is satisfied at relative order $m$.  For $m+k_1 < k_2$ $\phi(m+k_1)$ actually lies within
   $C^\infty(M,\setc)$. If, on the other hand, $\mu^2 \leq K$ and $m+k_1 = k_2$, then relation
   (\ref{sfg.noc}) implies the existence of a $\chi \in C^\infty(M,\setc)$ such that the
   scalar field equation is solved at relative order $m$ by setting $\phi(m+k_1):=\phi_0 +
   \chi Z$. This, in particular, suffices (P2).
 \end{description}
 Uniqueness up to but not including relative order $n$ follows with the same argument by taking
 advantage of the necessary conditions derived in section \ref{esfg}.
\end{proof}

The condition for the existence of a solution of the algebraic Einstein scalar field equations
is stated by means of a scalar $\zeta$ and a 1-form $Z_b$. For defining those, assume
$A_{ab}$ being a Riemannian metric and $\phi_0$, $\phi_1$ smooth complex valued functions
on $M$. By virtue of Lemma \ref{initsol} and equations (\ref{egevol.tl}), (\ref{egevol.tr}),
(\ref{cgevol.g}) there exists a solution $(g_{ab},\phi)$ of the Einstein scalar field equations
up to relative order $n$, exclusive, with (G1), (P1), (P2) valid and such that the coefficients
$(\hE\ud{a}{b})_{n,0,l}$ and $(E)_{n,0,l+1}$ of the evolution quantities vanish too for all
natural $l$. It can then be shown that the right hand sides of
\begin{eqnarray}
 \zeta(A,\phi_0,\phi_1) & := & A^{ij} (g_{ij})_{n-2,0,0} - \frac{2}{n^2 H^2} (E)_{n,0,0}
  \label{zeta} \\
 Z_b(A,\phi_0,\phi_1) & := & \ndel^i (g_{ib})_{n-2,0,0} - \frac{2}{n^2 H^2} \left( \ndel_b
  (E)_{n,0,0} + nH (C_b)_{n,0,0} \right) \hspace{2em} \label{Zeta}
\end{eqnarray}
are independent of the choice of such a solution $(g_{ab},\phi)$ and therefore the functions
$\zeta$ and $Z_b$ are well-defined. Now the necessary part of the condition is obvious: If
$(g_{ab},\phi)$ is a solution of the Einstein scalar field equations for which (G1), (P1) and (P2)
holds, then $A^{ij} (g_{ij})_{n-2,0,0} = \zeta(A,\phi_0,\phi_1)$ and $\ndel^i (g_{ib})_{n-2,0,0} =
Z_b(A,\phi_0,\phi_1)$. The sufficient part is contained in the next theorem.

\begin{theorem} \label{solution}
 Let $A_{ab} \in C^\infty(M)$ be the components of a Riemannian metric, $B_{ab} \in C^\infty(M)$
 those of a symmetric tensor and $\phi_0, \phi_1 \in C^\infty(M,\setc)$ smooth complex valued
 functions on $M$ with
 \begin{displaymath}
  A^{ij} B_{ij} = \zeta(A,\phi_0,\phi_1) \ {\rm and} \ \ndel^i B_{ib} = Z_b(A,\phi_0,\phi_1).
 \end{displaymath}
 Then there exists exactly one solution $(g_{ab}, \phi)$ of the Einstein scalar field equations
 with $g_{ab} \in \sS_n X^{-2}$, $\phi \in \sS_{k_2-k_1} X^{k_1}$ and such that
 \begin{displaymath}
  \begin{array}{llrcl}
   {\rm (G1)} & \hfill & (g_{ab})_{-2,0,0} & = & A_{ab} \\
   {\rm (G2)} & & (g_{ab})_{n-2,0,0} & = & B_{ab} \\[1ex]
   {\rm (P1)} & & \phi(k_1) & = & \left\{
   \begin{array}{lll}
    \phi_0 & {\rm for} & \mu^2 < K \\
    \phi_0 + \phi_1 Z & {\rm for} & \mu^2 = K \\
    \phi_0 Y^{-\omega} + \phi_1 Y^\omega & {\rm for} & \mu^2 > K
   \end{array} \right. \\[4ex]
   {\rm (P2)} & & \lefteqn{ \phi(k_2) \in \phi_1 + C^\infty(M,\setc)Z \ {\rm for} \ \mu^2 <
   K } \hfill
  \end{array}
 \end{displaymath}
 are valid altogether.
\end{theorem}

\begin{proof}
 The assertion is proved by induction over relative orders in $D$. By virtue of Lemma
 \ref{initsol} it suffices to consider relative orders greater than or equal to $n$.  So, for a $m
 \geq n$ let $g_{ab} : \ioo{-\infty,m-2} \to C^\infty(M)[Z]\py$ and $\phi : \ioo{-\infty,m+k_1}
 \to C^\infty(M,\setc)[Z]\py$ be constructed such that $(g_{ab})$ is real symmetric, the
 Einstein scalar field equations are satisfied up to relative order $m$, exclusive, and the
 conditions (G1), (P1), (P2), and for $m>n$ also (G2) hold. Moreover, the support of $g_{ab}$
 may be finite, having exactly the point $-2$ lying below $k_0-2$, the support of $\phi$ be
 finite as well, having no points lying below $k_1$. Finally, for all $t$ less than $n-2$,
 $g_{ab}(t)$ may belong to $C^\infty(M)$, as for $t < k_2$, $\phi(t)$ may be within the
 subalgebra $C^\infty(M,\setc)$. Then the metric is extended to relative order $m$ as follows:
 \begin{description}
  \item[$m=n:$]
   The equations (\ref{egevol.tl}), (\ref{egevol.tr}) together with (\ref{cgevol.g}) ensure the
   existence of real $g_{ab}(n-2) \in C^\infty(M,\setc)[Z]\py$ with $(g_{ab})_{n-2,0,0}=B_{ab}$
   such that $(g_{ab})$ is symmetric and causes $\hE\ud{a}{b}=0$ and $E(n) \in C^\infty(M,\setc)$,
   i.e. all but the coefficient $(E)_{n,0,0}$ necessarily vanish. Due to the trace condition
   implied on $B$, this coefficient turns out to be zero too, so $E(n)=0$ and by consistency
   equation (\ref{cc.ham}) $C(n)=0$ is obtained.  In the same manner, relation (\ref{cc.mom})
   firstly implies $C_a(n) \in C^\infty(M,\setc)$ and the divergence condition on $B$ then
   guarantees that $C_a(n)$ actually vanishes. So, the Einstein equations are satisfied at
   relative order $n$.
  \item[$m=2n:$]
   In this case, the relations (\ref{egevol.tl}), (\ref{egconstr.ham}) and (\ref{cgevol.g}) give a
   real symmetric family of $g_{ab}(m-2) \in C^\infty(M,\setc)[Z]\py$ so that $\hE\ud{a}{b}(m)=0$
   and $C(m)=0$ are fulfilled. The consistency conditions (\ref{cc.ham}) and (\ref{cc.mom})
   then show $E(m)=0$ and $C_a(m)=0$ successively and so the validity of the Einstein equations
   at relative order $m$.
  \item[$m \notin \{n,2n\}:$]
   Here, the relations (\ref{egevol.tl}), (\ref{egevol.tr}) and (\ref{cgevol.g}) yield a real
   symmetric family of $g_{ab}(m-2) \in C^\infty(M,\setc)[Z]\py$ such that $\hE\ud{a}{b}(m)=0$
   and $E(m)=0$ are satisfied. Once again, the consistency conditions (\ref{cc.ham}) and
   (\ref{cc.mom}) are used to show $C(m)=0$ and $C_a(m)=0$ successively and so the validity
   of the Einstein equations at relative order $m$.
 \end{description}
 The continuation of the scalar field can, in any case, be done using the relations
 (\ref{sfg.noc}) and (\ref{sfg.oc}), respectively, to provide a coefficient $\phi(m+k_1) \in
 C^\infty(M,\setc)[Z]\py$ such that the scalar field equation holds at relative order $m$ too.

 The uniqueness of such a solution follows with the same inductive argument from the
 presuppositions of the theorem by making use of the uniqueness of the initial solution as
 stated in Lemma \ref{initsol}.
\end{proof}

It is worth noting that for non-subcritical field masses the supports of the series become
especially simple because then $k_1=k_2=n/2$ and so the set $D$ consists of natural or halves
of natural numbers only, depending on whether $n$ is even or odd, respectively. Oscillatory
terms by their part occur exclusively at integer multiples of $\omega$. If the field mass is
not supercritical, there are no oscillatory terms at all.

The restriction to masses $\mu^2 > - \xi n (n+1) H^2$ and thus to fields of order greater than
zero is stringent for the treatment outlined in this section since just by allowing equality
the right hand side of (\ref{hubble}) becomes, in general, dependent on the field and so the
Einstein equations cannot necessarily be satisfied to relative order zero with a constant $H$.

%
%

\section{Discussion} \label{discussion}

In this paper, existence and uniqueness of formal power series solutions for complex scalar
field models with arbitrary coupling to the scalar curvature is proven for a large class of
self interaction potentials. The class includes potentials which, written as a function of the
absolute field amplitude, possess a positive minimum at zero. The algebraic analogues of the
Einstein scalar field equations were obtained by considering the model in $n+1$ decomposition
with respect to Gaussian coordinates and postulating those relations to hold on a suitable
algebra of formal series endowed with all the necessary operations. This yielded algebraic
equations for the series' coefficients which could be solved inductively. The series used here
are generalizations of those introduced by Rendall \cite{Ren0312} for a similar treatment of
the vacuum case and perfect fluid models since, apart from oscillatory terms, both logarithmic
terms as well as non-integer powers of the exponential factor $e^{-H t}$ may occur.

The lower mass limit that is imposed in the case of non-positive coupling constants can be
interpreted as a condition for the field to decay exponentially in the future. For a more
specific classification of the asymptotics, a critical field mass $\sqrt{K} = \sqrt{n^2/4 -
\xi n (n+1)} H$ is found that distinguishes solutions based on the occurrence of logarithmic
or oscillatory terms in the leading order $k_1$ of the field's expansion. For subcritical
masses, $\mu^2 < K$, neither logarithmic nor oscillatory terms are present in leading
order, whereas for critical masses $\mu^2=K$, the lowest logarithmic term can appear already
initially. Both correspond to fields that would fall off exponentially with increasing time and
this asymptotically at least like $e^{-k_1 H t}$ in the former, at least like $e^{-\beta H t}$
for all $\beta$ less than $k_1$ in the latter case. Thereby, the exponent $k_1$ increases from
arbitrary small positive values for the lowest permitted masses to $n/2$ for the critical
field mass. Supercritical masses on the other hand allow oscillatory terms in the series
to show up, at frequency $\pm \omega H = \pm \sqrt{\mu^2-K}$ in leading order $k_1=n/2$
and together with overtones thereof in higher orders. This has an interpretation as a field
that could oscillate asymptotically at frequency $\omega H$ and whose amplitude would decay
at least like $e^{-n H t / 2}$ for large times. The higher frequency oscillations would be
attenuated exponentially with respect to the amplitude of the first harmonic.

It should be emphasized that the connection between certain coefficients in the formal series
and the asymptotics of a scalar field as made above is just a prospect of the large time
behaviour the solutions might be expected to show. There is no assertion that the series
would in fact be asymptotic in the sense of theorem 5 of \cite{Ren0312}, or even convergent,
when $t$ is considered a time coordinate as in section \ref{motivation}. Drawing this analogy
nonetheless, the results obtained here are found to coincide with the asymptotics of scalar
field models derived analytically by Rendall \cite{Ren0403,Ren0408}. In particular, the
expression for the late time limit of the Hubble parameter (\ref{hubble}) or the asymptotic
value $-n H$ for the trace of the second fundamental form are identical. Additionally, the
condition on the second derivative of the potential necessary for an oscillating decay of the
field agrees with $\mu^2 > K$. It might also be interesting that recent SN Ia data suggests,
although not by itself significantly enough to rule out even $\Lambda$CDM, the existence of
oscillations in the cosmological scale factor \cite{Laz05,Luo05}. In \cite{Laz05}, Lazkoz et
al. achieved their best fit with a purely empirical oscillatory model. The observed decay of
the oscillation's amplitude with the third power of the scale factor is compatible with the
first possible occurrence of oscillatory terms in the expansion of the metric found here.

\vspace{1em}
Part of this work was done in the course of a diploma thesis written at the University of Berne,
Switzerland, supervised by Prof. Dr. P. H\'aj\'\i\v cek, in 2004.

%
%

\end{document}